\documentclass[cameraready]{Interspeech}
\usepackage{amsmath,graphicx,hyperref,makecell,arydshln}
\usepackage{siunitx}
\usepackage{comment}
\usepackage{booktabs}   
\usepackage{amssymb}

\usepackage{multirow}   
\usepackage{colortbl}   
\usepackage{xcolor}     
\usepackage{bm}
\usepackage[table]{xcolor}
\definecolor{lightgray}{gray}{0.93}
\title{Semantic-VAE: Semantic-Alignment Latent Representation for Better Speech Synthesis}

\author[affiliation={1,2}]{Zhikang}{Niu}
\author[affiliation={3}]{Shujie}{Hu}
\author[affiliation={4}]{Jeongsoo}{Choi}
\author[affiliation={1,2}]{Yushen}{Chen}
\author[affiliation={1}]{Peining}{Chen}
\author[affiliation={5}]{Pengcheng}{Zhu}
\author[affiliation={5}]{Yunting}{Yang}
\author[affiliation={5}]{Bowen}{Zhang}
\author[affiliation={5}]{Jian}{Zhao}
\author[affiliation={5}]{Chunhui}{Wang}
\author[affiliation={1,2}, correspondingauthor]{Xie}{Chen}

\address{
    $^1$ X-LANCE Lab, MoE Key Lab of Artificial Intelligence, Shanghai Jiao Tong University, China \\
    $^2$ Shanghai Innovation Institute, China; $^3$ The Chinese University of Hong Kong, China \\
    $^4$ School of Electrical Engineering, KAIST, South Korea; $^5$ Geely, China
}

\email{zhikangniu@sjtu.edu.cn, chenxie95@sjtu.edu.cn}


\begin{document}

\maketitle

\begin{abstract}
Mel-spectrograms have been widely used in zero-shot text-to-speech (TTS); their inherent redundancy leads to inefficiency in text-speech alignment. Compact VAE-based latent representations have emerged as a stronger alternative but exhibit an optimization dilemma: higher-dimensional latents improve reconstruction quality and speaker similarity but degrade intelligibility, while lower-dimensional latents improve intelligibility at the cost of reconstruction fidelity. To overcome this dilemma, we propose Semantic-VAE, which uses semantic alignment regularization in the latent space. This design alleviates the reconstruction-generation trade-off by capturing semantic structure in high-dimensional latent representations. 
When integrated into F5-TTS, our method achieves 2.10\% WER and 0.64 speaker similarity on LibriSpeech-PC, outperforming mel-based systems and vanilla acoustic VAE baselines with improved training efficiency.
Demo and codes: https://zhikangniu.github.io/semantic-vae/
\end{abstract}

\section{Introduction}
\label{sec:intro}
Zero-shot Text-to-Speech (TTS) aims to synthesize natural human speech from text inputs, conditioned on a reference speech prompt to closely mimic the speaker’s timbre. In recent years, zero-shot TTS models have achieved remarkable progress by scaling up both data and model size. Existing methods for zero-shot TTS can be broadly categorized into two approaches: Autoregressive (AR) ~\cite{wang23valle,anastassiou2024seed,peng24voicecraft,lajszczak2024base,meng2024autoregressive} and Non-Autoregressive (NAR)~\cite{le2024voicebox,Shen24naturalspeech,Ju24naturalspeech,chen2024f5,lee2024ditto,jiang2025megatts} models. 

Most AR models directly quantize speech into discrete tokens and achieve exceptional performance in capturing and reproducing the acoustic characteristics of a reference speaker through in-context learning. However, they suffer from slow inference speed and error accumulation due to autoregressive generation of speech representation and information loss from quantization~\cite{chen24valle2,han24valler,vallt25}. NAR models, primarily built on diffusion and flow matching models and using continuous speech representations, address the inherent limitations of AR models and overcome the information loss introduced by quantization, thereby improving both inference speed and output quality. NAR-based models can be grouped into two main categories based on their intermediate representation: 
\begin{figure}[tbp]
  \centering
  \includegraphics[width=7cm]{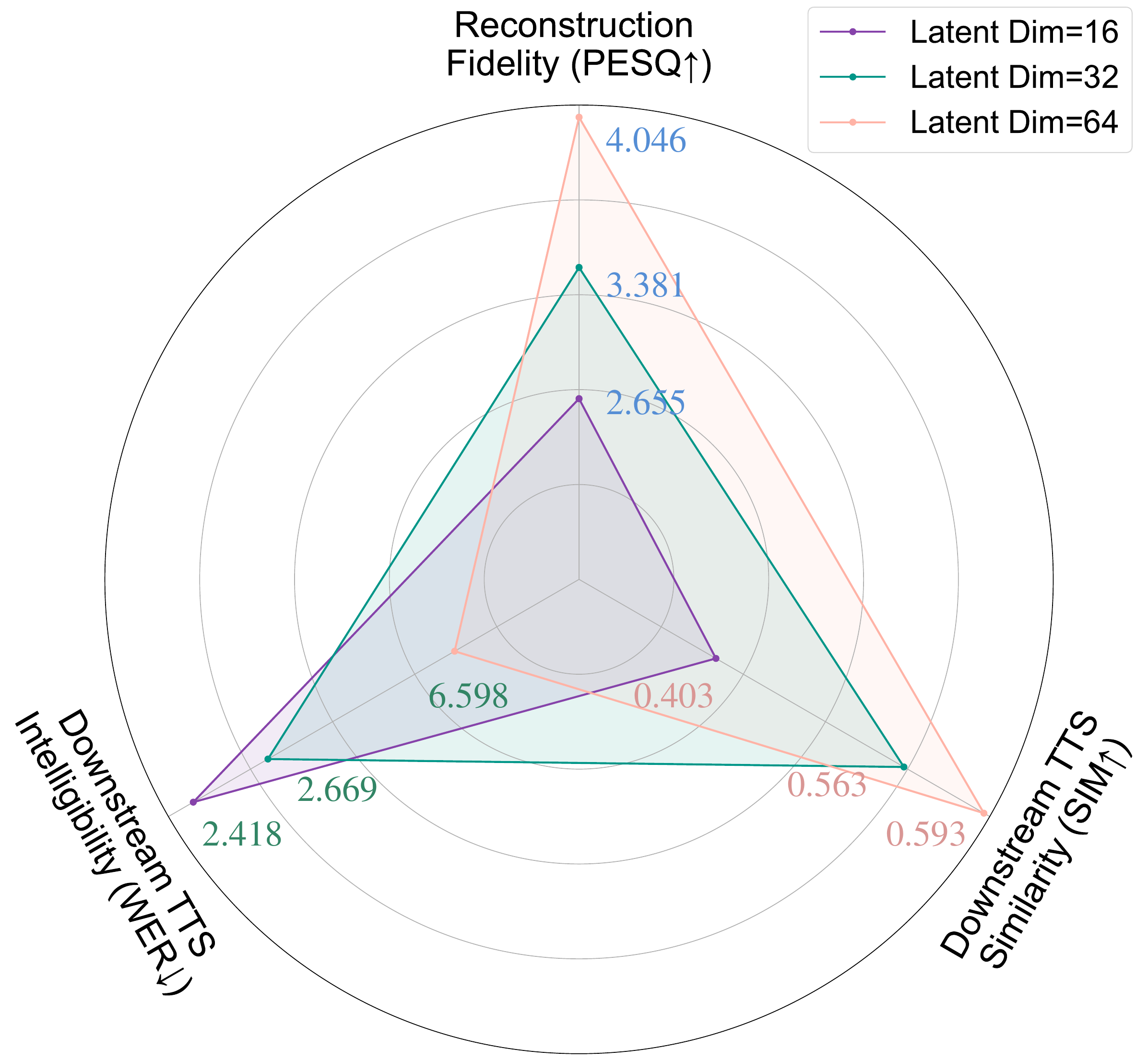}
  \caption{Reconstruction and Generation(TTS) dilemma on continuous representation. Higher-dimensional latents retain more acoustic information, improving reconstruction and speaker similarity; lower-dimensional latents focus on semantic information, boosting intelligibility while trading off reconstruction and similarity.
  }
  \vspace{-0.5cm}
  \label{fig:dilemma}
\end{figure}

\par
\noindent\textbf{1) Models conditioned on the mel-spectrogram}, which constitute the predominant paradigm in NAR-based TTS, use a pre-trained vocoder to reconstruct the waveform from the generated mel-spectrogram. Representative examples include VoiceBox \cite{le2024voicebox}, E2 TTS \cite{eskimez2024e2}, and F5-TTS \cite{chen2024f5}. Despite the strong performance of mel-based systems, they still suffer from several inherent limitations \cite{liu2022delightfultts,lee2025wave,qiang2024high}, including loss of phase information, high dimensionality with substantial redundancy, and the absence of fine-grained high-frequency details, which limit high-fidelity speech synthesis.
\par
\noindent\textbf{2) Models conditioned on latent representations} produced by a Variational Autoencoder (VAE) and synthesize waveforms directly using the corresponding decoder. VAE-based systems mitigate the drawbacks mentioned above by learning compact latent representations that preserve essential acoustic information while eliminating redundancy. These compact representations not only accelerate training convergence but also yield improved synthesis quality~\cite{jiang2025megatts}. For instance, $\text{SeedTTS}_{DiT}$~\cite{anastassiou2024seed} is a large-scale state-of-the-art industrial TTS model that leverages VAE latent representations, highlighting the effectiveness of compact representation for high-fidelity speech synthesis. Building on this idea, MegaTTS-3~\cite{jiang2025megatts} further introduces a novel sparse alignment strategy to guide a latent diffusion transformer over the VAE latent space, achieving performance gains. 

Consistent with findings in the vision domain \cite{yao2025reconstruction}, latent diffusion models with vanilla acoustic VAE systems face an \textbf{optimization dilemma} between reconstruction and generation performance. The results of preliminary experiments are shown in Fig. \ref{fig:dilemma}. Specifically, increasing the latent dimension enhances reconstruction quality and speaker similarity but impairs intelligibility in zero-shot TTS, whereas reducing the latent dimension improves intelligibility at the expense of reconstruction quality and speaker similarity. This trade-off exemplifies the information bottleneck: lower-dimensional representations capture core semantic content, whereas higher-dimensional representations preserve richer acoustic details but introduce redundancy and complicate semantic modeling.

To address this challenge, we introduce \textbf{Semantic-VAE}, a novel VAE with semantic alignment regularization in the latent space to mitigate the difficulty of semantic modeling in zero-shot TTS tasks with unconstrained high-dimensional latent representations. Extensive experiments demonstrate that Semantic-VAE mitigates the generation–reconstruction dilemma in high-dimensional latent spaces, accelerates training convergence, and achieves state-of-the-art performance in zero-shot TTS tasks. Specifically, the F5-TTS model with the proposed Semantic-VAE features achieves \textbf{2.10}\% WER and \textbf{0.64} speaker similarity on the LibriSpeech-PC test-clean dataset, outperforming the mel-based F5-TTS baseline (2.23\%, 0.60) and the vanilla acoustic VAE baseline (2.65\%, 0.59). Additional reconstruction experiments show that Semantic-VAE achieves reconstruction quality comparable to vanilla VAE and substantially better than the vocoder~\cite{Siuzdak24vocos} with mel-spectrograms. This demonstrates that our proposed semantic alignment regularization enables more effective generative modeling without compromising the representational capacity of the latent features. 

\section{Method}
\label{sec:method}
In this section, we provide a detailed introduction of our proposed method, highlighting its key components and underlying motivations. Section \ref{sec:vae} introduces our baseline VAE architecture and objective. Section \ref{sec:align} presents our strategy for semantic regularizing latent representations using pre-trained self-supervised learning (SSL) speech models. 
\begin{figure}[htbp]
  \centering
  \includegraphics[width=6cm]{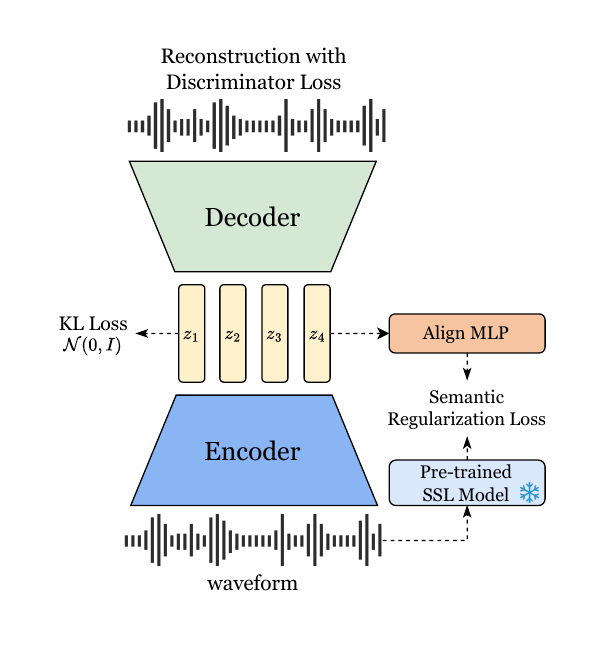}
  \caption{Overall training framework. The encoder maps the input into latent representations, regularized by a KL loss.
  The decoder reconstructs the waveform guided by a discriminator loss, while a semantic regularization loss aligns the latent space with features from a frozen pre-trained SSL model.
  }
  \label{fig:1}
\end{figure}

\subsection{Variational Autoencoder}
\label{sec:vae}
To obtain latent representations for speech generation, we employ a Variational Autoencoder (VAE) \cite{kingma2014auto} that models speech waveforms.
The VAE consists of an encoder that maps the input speech $\bm{x}$ into a latent representation $\bm z$, and a decoder that reconstructs the signal $\hat{\bm x}$ from $\bm z$. The standard training objective of a VAE maximizes the Evidence Lower Bound (ELBO):
\begin{equation}
\resizebox{0.91\columnwidth}{!}{$
\mathrm{ELBO} :=
\mathbb{E}_{q_\phi(\mathbf{z}\mid \mathbf{x})}\bigl[\log p_\psi(\mathbf{x}\mid \mathbf{z})\bigr]
- \mathcal{D}_{\mathrm{KL}}\bigl(q_\phi(\mathbf{z}\mid \mathbf{x})\parallel p(\mathbf{z})\bigr)
$}
\end{equation}
where $q_\phi(\bm z|\bm x)$ denotes the variational posterior that approximates the true latent distribution, $p_\psi(\bm x|\bm z)$ models the conditional likelihood of the reconstructed speech $\hat{\bm x}$ given $\bm z$, and the KL divergence term regularizes the latent space towards the standard Gaussian prior $p(\bm z)=\mathcal{N}(0,\bm I)$.

Following the standard VAE objective, we express the generative loss as a weighted sum of the reconstruction loss $\mathcal{L}_{\text{recon}}$ and the KL divergence loss $\mathcal{L}_{\text{KL}}$:
\begin{equation}
\mathcal{L}_{\text{gen}} = 
\lambda_\text{recon}\mathcal{L}_{\text{recon}} + \lambda_\text{KL}\mathcal{L}_{\text{KL}},
\end{equation}
where $\lambda_\text{KL}$ balances the trade-off between reconstruction fidelity and latent space regularization. Following DAC~\cite{kumar2023high}, we use multi-scale frequency domain reconstruction loss $\mathcal{L}_{\text{recon}}$, defined as L1 distance between mel-spectrograms of the ground-truth and the reconstruction over multiple window lengths.

Moreover, to enhance the perceptual quality of the output, we adopt adversarial training with a multi-period discriminator~\cite{kong2020hifi} and a multi-band, multi-scale STFT discriminator~\cite{kumar2023high}. We use a HingeGAN~\cite{lim2017geometric} adversarial objective with L1 feature-matching loss.
The overall VAE training objective is then expressed as follows:
\begin{equation}
\mathcal{L}_{\text{VAE}}
= \mathcal{L}_{\text{gen}}
+ \lambda_{\text{adv}}\,\mathcal{L}_{\text{adv}}
+ \lambda_{\text{feat}}\,\mathcal{L}_{\text{feat}}.
\end{equation}

\subsection{Semantic Representation Alignment}
\label{sec:align}

The reconstruction and generation dilemma is a widely discussed challenge in discrete audio codecs. Previous studies~\cite{zhang2024speechtokenizer, Ju24naturalspeech} have demonstrated that aligning with pre-trained SSL speech models (e.g., HuBERT~\cite{hsu2021hubert}, WavLM~\cite{chen2022wavlm}, w2v-BERT~\cite{chung2021w2v}, etc.), which capture rich phonetic and semantic representations, can provide more informative supervision for audio codec training and improve downstream AR model performance and convergence. Similarly, incorporating SSL alignment into Diffusion Transformers (DiT) training accelerates convergence and yields better performance in speech synthesis~\cite{choi2025accelerating} and image generation~\cite{yu2024representation}.

However, to the best of our knowledge, no prior work has explored applying semantic regularization to speech continuous latent spaces, nor examined its potential to improve downstream task performance. Motivated by these insights, we introduce a semantic regularization loss into the VAE training framework. This addition aims to investigate whether semantic-aligned VAEs can improve both the convergence speed and performance of NAR-based TTS models compared to vanilla acoustic VAEs. Specifically, we employ a pre-trained SSL speech model $f(\cdot)$ to extract structured semantic representations from the same speech $\bm x$. The hidden states $\bm h$ of the SSL model, aligned in both temporal and feature dimensions with the VAE latent representation, are obtained through an interpolation layer followed by a 1D convolution layer, formulated as
$\bm h=\text{Conv1D}(\text{Interp}(f(\bm x))).$
The semantic alignment regularization loss is then defined as
\begin{equation}
    \mathcal{L}_{\text{Align}} = -\frac{1}{T} \sum_{t=1}^{T} \cos\big(\bm h^{[t]}, \, \bm z^{[t]} \big),
\end{equation}
where $T$ denotes the sequence length and $\text{cos}(\cdot,\cdot)$ computes the cosine similarity between the aligned VAE latent and the SSL feature at each time step $t$. This loss regularizes the VAE latent space, encouraging the VAE to learn a structured semantic representation from SSL models. To jointly optimize for high-fidelity speech reconstruction and semantically meaningful latent representations, we integrate the VAE training objective with the proposed semantic regularization loss. The overall training objective is formulated as  
\begin{equation}
\mathcal{L}_{\text{total}}
= \mathcal{L}_{\text{VAE}}
+ \lambda_{\text{Align}} \, \mathcal{L}_{\text{Align}},
\end{equation}
where $\lambda_\text{Align}$ is a hyperparameter that controls the relative importance of the semantic alignment regularization. 

\section{Experimental Setup}
\label{sec:exp}
In this section, we describe implementation details, datasets, and the evaluation setup.
\subsection{Semantic-VAE Model Setup}
Following DAC~\cite{kumar2023high}, $\text{Semantic-VAE}$ adopts the same convolutional encoder, discriminator, and loss weights. 
To improve reconstruction performance, we replace the original convolutional decoder with an AMP Block-based decoder~\cite{lee2022bigvgan}. The encoder downsamples the 16 kHz input $\bm x$ with factors of $[4, 4, 5, 5]$, producing a 64-dimensional latent representation $\bm z$ at a frame rate of 40 Hz, while the decoder upsamples $\bm z$ with factors of $[5,5,2,2,2,2]$ to reconstruct the waveform.

$\text{Semantic-VAE}$ is trained for 600k iterations with a global batch size of 64. Each sample is a 3-second segment and resampled to 16 kHz, totaling approximately 192 seconds per batch. We employ the Adam optimizer with a learning rate of \(1 \times 10^{-4}\) and apply exponential decay with a factor of $\gamma = 0.9996$. In our experiment, we set $\lambda_\text{Align}=1$, $\lambda_\text{KL}=0.01$, $\lambda_\text{adv}=1$, $\lambda_\text{feat}=2$, and $\lambda_\text{recon}=15$.
\subsection{Zero-shot TTS Model Setup}
We use the F5-TTS~\cite{chen2024f5} as the baseline and adopt its training setup. In our approach, the mel-spectrograms are replaced with VAE latent representations. The model is optimized using the AdamW optimizer with \(7.5 \times 10^{-5}\), following a warm-up schedule for the first 20,000 updates and linear decay thereafter. For inference, we follow the same setup as F5-TTS, applying the sway sampling strategy and using the Euler ODE solver.
\subsection{Training Datasets}
We train Semantic-VAE on LibriTTS~\cite{Zen19libritts} and the small/medium subsets of Libriheavy~\cite{kang2023libriheavy}, totaling 6k hours. Unless otherwise stated, 
TTS models are only trained on LibriTTS.
\subsection{Evaluation}
To comprehensively evaluate Semantic-VAE regarding the optimization dilemma between reconstruction and generation, we consider two tasks: \textbf{1) reconstruction} and \textbf{2) downstream zero-shot TTS}. For the reconstruction task, we use the LibriTTS test-other as the test set, and employ UTMOS~\cite{Saeki22UTMOS}, PESQ~\cite{pesq_rix}, and STOI as the evaluation metrics to quantify perceptual quality and intelligibility. For the zero-shot TTS evaluation, we follow the setup in F5-TTS\footnote{\href{https://github.com/SWivid/F5-TTS/tree/main/src/f5_tts/eval}{https://github.com/SWivid/F5-TTS/tree/main/src/f5\_tts/eval}} and evaluate on the LibriSpeech-PC~\cite{meister2023librispeech} test-clean. We perform the cross-sentence generation and evaluate the performance using the Word Error Rate (WER), Speaker Similarity (SIM-o), and UTMOS, which respectively measure intelligibility, speaker consistency, and naturalness. We further conduct 5-point MOS listening tests with 25 listeners (250 ratings per system).

\begin{table}[t]
  \renewcommand{\arraystretch}{1.3}
  \renewcommand{\tabcolsep}{2.3mm}
  \centering
  \caption{Zero-shot TTS results on the LibriSpeech-PC test-clean. The best results in the low-resource setting are marked in \textbf{bold}. High-resource results are quoted from the F5-TTS for comparison.}
  \label{tab:table1}
  \resizebox{0.999\linewidth}{!}{
  \begin{tabular}{l c c c c c}
    \Xhline{3\arrayrulewidth}
    \textbf{Method} & \# \textbf{Param.} & \textbf{Data} & \textbf{WER(\%)$\downarrow$} & \textbf{SIM$\uparrow$} & \textbf{MOS$\uparrow$} \\
    \hline
    Ground Truth & - & - & 2.23 & 0.69 & 4.20 $\pm$ 0.08 \\
    \hdashline
    \multicolumn{6}{c}{\itshape High-resource} \\
    CosyVoice~\cite{Du24cosyvoice} & 300M & 170kh & 3.59 & 0.66 & - \\
    FireRedTTS~\cite{guo2024fireredtts} & 580M & 248kh & 2.69 & 0.47 & - \\
    E2 TTS~\cite{eskimez2024e2} & 333M & 100kh & 2.95 & 0.69 & - \\
    F5-TTS~\cite{chen2024f5} & 336M & 100kh & 2.42 & 0.66 & - \\
    \hdashline
    \multicolumn{6}{c}{\itshape Low-resource} \\
    USLM~\cite{zhang2024speechtokenizer} & 361M & 0.6kh & 6.11 & 0.43 & - \\
    E2 TTS~\cite{eskimez2024e2} & 157M & 0.6kh & 3.51 & 0.61 & 3.68 $\pm$ 0.11 \\
    ~~~+ \textbf{Semantic-VAE }& 157M & 0.6kh & 2.41 & 0.62  & 3.88 $\pm$ 0.10 \\
    F5-TTS~\cite{chen2024f5} & 159M & 0.6kh & 2.23 & 0.60 & 3.99 $\pm$ 0.10 \\
    ~~~+ Vanilla VAE & 159M & 0.6kh & 2.65 & 0.60 & 4.13 $\pm$ 0.09 \\
    ~~~+ \textbf{Semantic-VAE} & 159M & 0.6kh & \textbf{2.10} & \textbf{0.64} & \textbf{4.17 $\pm$ 0.09} \\
    \Xhline{3\arrayrulewidth}
  \end{tabular}}
  \vspace{-0.2cm}
\end{table}
\section{Results}
\subsection{Downstream Zero-shot TTS Results}
\label{tts_results}
\begin{figure}[ht]
    \centering
    \includegraphics[width=8cm]{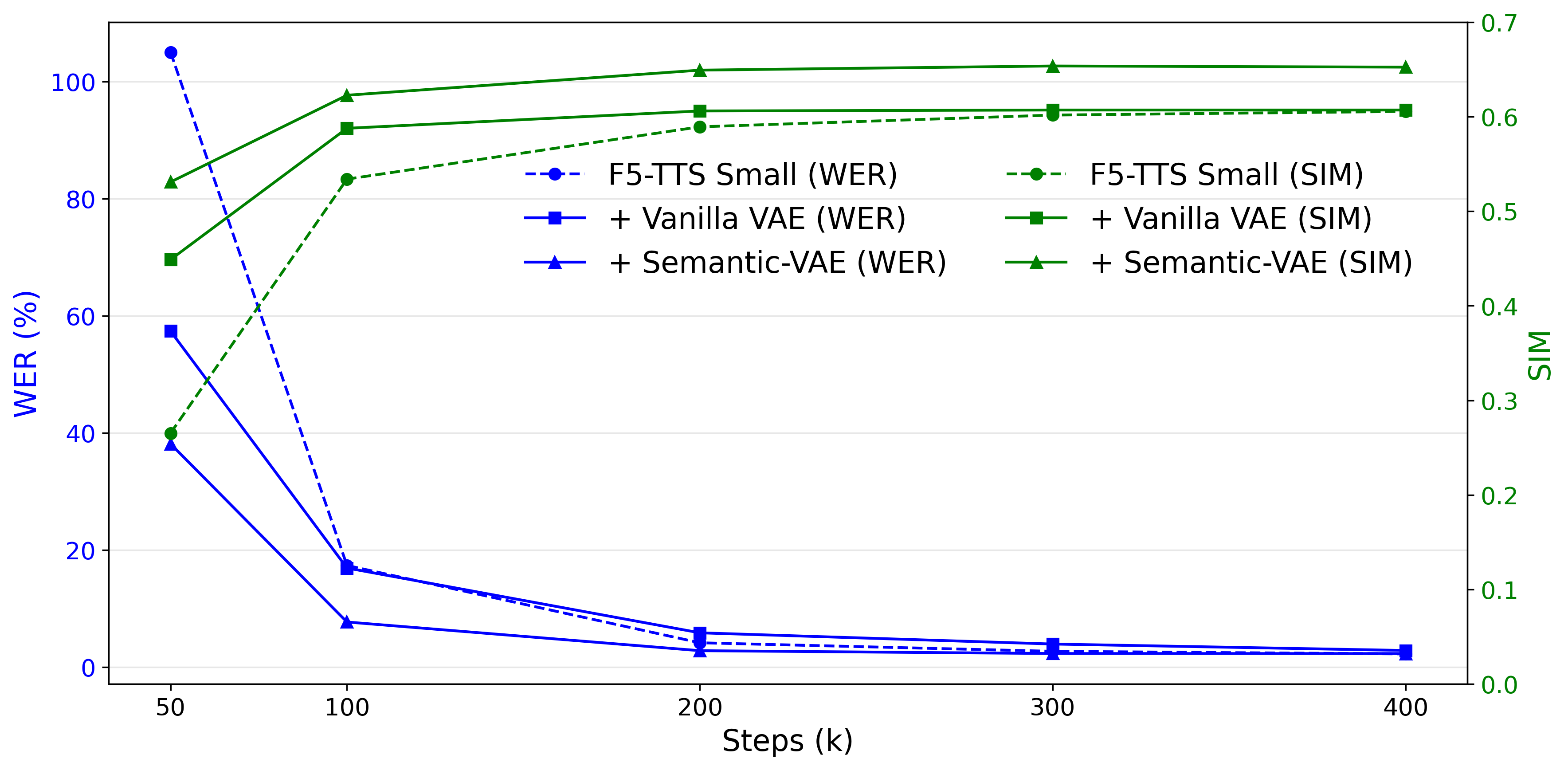}
    \caption{Comparison of WER and SIM on the LibriSpeech-PC test-clean dataset between the F5-TTS baseline, Vallina VAE, and Semantic-VAE across different training steps.}
    \label{fig:2}
    \vspace{-0.2cm}
\end{figure}
Table~\ref{tab:table1} reports zero-shot TTS results on LibriSpeech-PC test-clean. We include high-resource results from prior work for reference and focus our evaluation on the low-resource setting.
When integrated into F5-TTS, Semantic-VAE achieves the best WER and SIM among all low-resource models, outperforming both the F5-TTS baseline and its vanilla VAE variant.
A similar trend is observed when applying Semantic-VAE to E2 TTS, yielding consistent gains over the baseline. To further assess convergence, we compare the proposed Semantic-VAE against the vanilla VAE and the mel-based F5-TTS baseline across different training steps.  As shown in Fig. \ref{fig:2}, Semantic-VAE converges faster and achieves better performance than both the vanilla VAE and the mel-based baseline at the same step. These results demonstrate that Semantic-VAE not only improves the final performance, especially speaker similarity, but also accelerates convergence and reduces training costs compared with the baseline F5-TTS.

\subsection{Scaling to a High-Resource TTS Results}
\label{scaling}
To further validate the effectiveness of Semantic-VAE, we conduct additional experiments using larger-scale downstream TTS models and training data.
Specifically, we trained the F5-TTS Base (336M parameters) integrated with Semantic-VAE on Emilia~\cite{emilia}(100k hours) and evaluated it on multiple test sets under the same training budget (400k updates). 
The results show consistent improvements in WER, SIM and UTMOS on LibriSpeech-PC and SeedTTS-eval~\cite{anastassiou2024seed}.
Notably, although Semantic-VAE is trained only on English data, it also produce improvements over the baseline on the Chinese test set, suggesting robust cross-lingual generalization.
\begin{table}[ht]
  \centering
  \caption{Scaling model and training data results under the same training budget (400k updates) for fair comparison. $^\dagger$ denotes results reported in the F5-TTS paper (1250k updates).}
  \label{tab:cross_dataset}
  \resizebox{\columnwidth}{!}{
  \renewcommand{\arraystretch}{1.15}
  \setlength{\tabcolsep}{1.5mm}
  \begin{tabular}{l l c c c}
    \toprule
    \textbf{Dataset} & \textbf{Method} & \textbf{WER$\downarrow$} & \textbf{SIM$\uparrow$} & \textbf{UTMOS$\uparrow$} \\
    \midrule
    \multirow{3}{*}{LibriSpeech-PC}
      & F5-TTS$^\dagger$ & 2.42 & 0.66 & 3.89 \\
      & F5-TTS(reproduce) & 2.34 & 0.63 & 3.92\\
      &\textbf{ + Semantic-VAE} & \textbf{2.04} & \textbf{0.71} & \textbf{4.38} \\
    \midrule
    \multirow{3}{*}{SeedTTS-en}
      & F5-TTS$^\dagger$ & 1.83 & 0.67 & 3.76\\
      & F5-TTS(reproduce) & 1.70 & 0.63 & 3.76\\
      & \textbf{+ Semantic-VAE} & \textbf{1.61} & \textbf{0.72} & \textbf{4.11}\\
    \midrule
    \multirow{3}{*}{SeedTTS-zh}
      & F5-TTS$^\dagger$ & 1.56 & 0.76 & 2.95\\
      & F5-TTS(reproduce) & 1.60 & 0.74 & 3.04\\
      & \textbf{+ Semantic-VAE} & \textbf{1.48} & \textbf{0.77} & \textbf{3.25}\\
    \bottomrule
  \end{tabular}
  }
  \vspace{-0.2cm}
\end{table}

\subsection{Reconstruction Results}
Table \ref{table:recon} reports reconstruction results on LibriTTS test-other. Despite its aggressively compact representation in both temporal and latent dimensions, vanilla VAE outperforms Vocos in reconstruction. It indicates that end-to-end VAE training encourages the latent to retain essential information while discarding redundancy. Moreover, Semantic-VAE maintains reconstruction performance comparable to vanilla VAE, demonstrating that semantic alignment simplified generative modeling without sacrificing reconstruction performance.
\begin{table}[ht]
 \renewcommand{\arraystretch}{1.3}
 \vspace{-6pt}
 \centering
 \caption{Reconstruction performance on LibriTTS test-other}
 \label{table:recon}
 \resizebox{\columnwidth}{!}{%
 \begin{tabular}{c c c c c c}
 \Xhline{3\arrayrulewidth}
   \textbf{Model} & \textbf{frame/s} & \textbf{dim} & \textbf{PESQ}$\uparrow$ & \textbf{STOI}$\uparrow$ & \textbf{UTMOS}$\uparrow$ \\
  \midrule
  Ground Truth & - & - & - & - & 3.48 \\
  \hdashline
   Vocos & 93.75 & 100 & 3.57 & 0.96 & 3.24 \\
   Vanilla VAE & 40 & 64 & \textbf{3.75} & \textbf{0.97} & \textbf{3.57} \\
   Semantic-VAE & 40 & 64 & 3.74 & 0.96 & 3.56 \\
 \Xhline{3\arrayrulewidth}
 \end{tabular}
  } 
  \vspace{-0.4cm}
\end{table}

\subsection{Ablation Study}
\begin{table}[!htbp]
 \renewcommand{\arraystretch}{1.3}
 \centering
 \caption{Effect of different SSL models, layer, and alignment methods on LibriSpeech-PC TTS test set. ``Avg." = average of all layers; ``Last" = final layer output.}
 \label{tab:table3}
 \resizebox{\columnwidth}{!}{
 \begin{tabular}{c c c c c}
 \Xhline{3\arrayrulewidth}
   \textbf{Align method}& \textbf{Layer} & \textbf{WER (\%)}$\downarrow$ & \textbf{SIM}$\uparrow$ & \textbf{UTMOS}$\uparrow$ \\
  \midrule
  Baseline & - & 2.65 & 0.59 & 4.42 \\
   \hdashline
$|\Phi_n - f(x)|$ & \multirow{3}{*}{\makecell[c]{WavLM \\ 23rd}} & 3.12 & 0.47 & 3.29 \\
$\lVert \Phi_n - f(x) \rVert_2$ &  & 4.37 & 0.48 & 4.16 \\
$-\cos(\Phi_n, f(x))$ &  & \textbf{2.10} & \textbf{0.64} & \textbf{4.39} \\
  \midrule
   \textbf{SSL Models} & \textbf{Layer} & \textbf{WER (\%)}$\downarrow$ & \textbf{SIM}$\uparrow$ & \textbf{UTMOS}$\uparrow$ \\
   \midrule
  \multirow{3}{*}{HuBERT}
 & 23 & 2.28 & 0.63 & 4.38 \\
 & Last & 2.33 & 0.50 & 4.41 \\
 & Avg. & 2.29 & 0.61 & 4.39 \\
 \hdashline
 \multirow{3}{*}{WavLM}
 & 23 & \textbf{2.10} & \textbf{0.64} & 4.39 \\
 & Last & 2.62 & 0.58 & 4.34 \\
 & Avg. & 2.31 & 0.63 & \textbf{4.42} \\
 \Xhline{3\arrayrulewidth}
 \end{tabular}
  } 
  \vspace{-0.2cm}
\end{table}

Table \ref{tab:table3} presents a comprehensive ablation study on semantic alignment methods and semantic guidance selection. The results indicate that negative cosine similarity loss provides a more effective alignment mechanism than the L1 or MSE losses. This can be intuitively explained that L1 and MSE overly constrain absolute numerical differences between the latent space and the semantic guidance, whereas cosine loss focuses on directional alignment in the latent space, which better captures structured semantic representations.

Moreover, we explore \textit{which semantic guidance is most effective for semantic alignment}. It observes that WavLM\cite{chen2022wavlm} generally outperforms HuBERT\cite{hsu2021hubert} on SIM and comparable performance on WER, which can be attributed to its speaker-aware training objectives. Moreover, we compare different SSL layers and find that while all SSL layers help reduce WER, their impact on SIM varies considerably. In particular, the final layer consistently leads to a substantial drop in SIM performance, likely due to distributional adjustments required for aligning with the SSL training target, thereby discarding speaker-specific cues. Averaging SSL features across all layers improves SIM compared to the last-layer feature, which suggests that averaging all layers introduces more information that can improve acoustic details. However, incorporating too much information may also introduce redundancy, which could degrade semantic modeling. To verify this hypothesis, we chose the 23rd layer of WavLM for supervision. In SUPERB\cite{yang2021superb}, this layer has the highest weight for WER and provides richer semantic representations.
Experimental results confirm that this layer offers an optimal trade-off between WER and SIM, providing practical guidance for selecting pre-trained SSL features for alignment.
\section{Conclusion}
\label{sec:conclusion}
In this paper, we investigate the optimization dilemma in high-dimensional latent spaces and propose Semantic-VAE, a speech variational autoencoder with semantic alignment regularization. By learning structured and semantically aligned latent representations, Semantic-VAE alleviates the reconstruction–generation dilemma and improves downstream zero-shot TTS performance. Experiments demonstrate that our approach accelerates convergence and achieves state-of-the-art results on the zero-shot TTS task. In addition, comprehensive analyses and ablation studies confirm the effectiveness of each proposed component. We believe that these promising findings pave the way for future research on advanced modality alignment techniques and further optimization of latent diffusion-based text-to-speech systems.
\newpage
\section{Acknowledgements}
This work was supported by the National Natural Science Foundation of China  (No. U23B2018), Shanghai Municipal Science and Technology Major Project under Grant 2021SHZDZX0102 and Yangtze River Delta Science and Technology Innovation Community Joint Research Project (2024CSJGG1100).
\section{Generative AI Use Disclosure}
Generative AI was used only for editing and polishing purposes and was not used to generate any substantial sections of the manuscript.
\bibliographystyle{IEEEtran}
\bibliography{mybib}

\end{document}